\documentclass{aa}     

\usepackage{txfonts}
\usepackage{graphicx}
\usepackage{amssymb}

\usepackage{natbib}

\newcommand{\eqn} [1] {
\begin{equation}
#1
\end{equation}}

\def\vsini {{v\!\sin\!i}}
\def\kms {\rm{km}\,\rm{s}^{-1}}

\def\teff {\mathrm{T}_{\mathrm{eff}}}
\def\msol {{\mathrm{M}_{\odot}}}
\def\rsol {{\rm{R}_{\odot}}}
\def\lsol {{\mathrm{L}_{\odot}}}

\def\logg {\log g}

\def\muHz {\mu\mbox{Hz}}
\def\cd   {cd$^{-1}$}

\def\gdor {$\gamma$ Dor}
\def\ds {$\delta$ Scuti}
\def\dss {$\delta$ Scuti stars}

\begin{document}

   \title{Modelling of the fast rotating $\delta$ Scuti star Altair}
   \authorrunning{Su\'arez, Buzasi \& Bruntt.}
   
   \author{J.C. Su\'arez,\inst{1}\thanks{also associated researcher at  LESIA, 
                                       Observatoire de Paris-Meudon, UMR8109}, 
           H. Bruntt\inst{2,3}, D. Buzasi\inst{2}}

   \offprints{J.C. Su\'arez}

   \institute{Instituto de Astrof\'{\i}sica de Andaluc\'{\i}a (CSIC), Granada, 
              Spain \\
              \email{jcsuarez@iaa.es}   
         \and Department of Physics, US Air Force Academy, 2354 Fairchild Dr., 
	      Ste. 2A31, USAF Academy, CO\\
	      \email{derek.buzasi@usafa.af.mil}
	 \and Copenhagen University, Astronomical Observatory,
              Juliane Maries Vej 30, DK-2100 Copenhagen \O.\\
	      \email{bruntt@phys.au.dk}}

   \date{Received ... / Accepted ...}

   \abstract{We present an asteroseismic study of the fast rotating star
             \object{HD\,187642} (\object{Altair}), recently discovered
	     to be a \ds\ pulsator.
	     We have computed models taking into account rotation
	     for increasing rotational velocities. We investigate the relation
	     between the fundamental
	     radial mode and the first overtone in the framework
	     of Petersen diagrams. The effects of rotation on such diagrams, which 
	     become important at rotational velocities above $150\,\kms$,
	     as well as the domain of validity of our seismic tools are discussed.
	     We also investigate the radial and non-radial modes 
	     in order to constrain models fitting the five most dominant observed
	     oscillation modes.	    	     
	    
            }

\maketitle


\section{Introduction\label{sec:intro}}

The bright star Altair ($\alpha$ Aql) was recently observed by 
\citet{Buzasi05} (hereafter paper-I) with the star tracker on the Wide-Field Infrared 
Explorer (WIRE) satellite. The overall observations span from  18 October until 
12 November 1999, with exposures of 0.5 seconds taken during around 40\% of the
spacecraft orbital period (96 minutes). The analysis of the observations made in 
paper-I reveals Altair 
to be a low-amplitude variable star ($\Delta\,m<1\,\mathrm{ppt}$), 
pulsating with at least 7 oscillation modes. These results suggest that 
many other non-variable stars may indeed turn out to be variable when
investigated with accurate space observations.

Since Altair lies toward the low-mass end 
of the instability strip and no abundance anomalies or Pop~II characteristics are shown, the
authors identified it as a \ds\ star.
The \dss\ are representative of intermediate mass stars with spectral types from A
to F. They are located on and just off the main sequence, in the faint part of 
the Cepheid instability strip (luminosity classes V \& IV). 
Hydrodynamical processes occurring
in stellar interiors remain poorly understood. \dss\ seem particularly suitable 
for the study of such physical process, eg.~(a) convective overshoot from the core, which 
causes extension of 
the mixed region beyond the edge of the core as defined by the Schwarzschild 
criterion, affecting evolution; and (b) the balance between meridional circulation 
and rotationally induced turbulence generates chemical mixing and 
angular momentum redistribution \citep{Zahn92}. 

From the observational side, great efforts have been made within last decades 
in developing the seismology of \dss\ within coordinated networks, e.g.: 
\emph{STEPHI} \citep{Michel00stephi} or \emph{DSN} \citep{Breger00,Handler00}. 
However, several aspects of 
the pulsating behaviour of these stars are not completely understood 
\citep[see][]{Templeton97}. For more details, an interesting review of unsolved 
problems in stellar pulsation physics is given in \cite{Cox02}. Due to the
complexity of the oscillation spectra of \dss, the identification
of detected modes is often difficult and require additional information
\citep[see for instance][]{Viskum98, Breger99}. A unique mode identification is
often impossible and this hampers the seismology studies for 
these stars. Additional uncertainties arise from the effect of rapid rotation, 
both directly, on the hydrostatic balance in the star and, perhaps more 
importantly, through mixing caused by circulation or instabilities induced by 
rotation. 

Intermediate mass stars, like A type stars (\dss, \gdor, etc) are known to be
rapid rotators. Stars with $100<\vsini<200\,\kms$ are no longer 
spherically symmetric but are oblate spheroids due to the centrifugal force. 
Rotation modifies the structure of the star and thereby the propagation 
cavity of the modes. The characteristic pattern of symmetric multiplets split by 
rotation is thus broken.
\begin{table}
   \begin{center}
    \caption{Altair observed frequencies (from paper I). From left to
             right, columns provide frequencies in \cd, in $\muHz$, 
	     amplitudes in ppm, and finally,
	     $f_1/f_{j=2,...,7}$ frequency ratios.}
    \vspace{1em}
    \renewcommand{\arraystretch}{1.2}
    \begin{tabular}[ht!]{ccccc}
    \hline
        & (c/d) & ($\muHz$)  & $A$ (ppm)  &  $f_1/f_{j\neq1}$  \\
    \hline
  $f_1$ &    15.768   &      182.50      &    413     &         \\
  $f_2$ &    20.785   &      240.56      &    373     &  0.759  \\
  $f_3$ &    25.952   &      300.37      &    244     &  0.607  \\
  $f_4$ &    15.990   &      185.07      &    225     &  0.986  \\
  $f_5$ &    16.182   &      187.29      &    139     &  0.974  \\
  $f_6$ &    23.279   &      269.43      &    111     &  0.677  \\
  $f_7$ &    28.408   &      328.80      &    132     &  0.555  \\
    \hline
  \end{tabular}
    \label{tab:freqobs}
   \end{center}
\end{table}
In the framework of a linear perturbation analysis, the second order effects 
induce strong asymmetries in the splitting of multiplets \citep{Saio81,DG92} 
and shifts which cannot be neglected even for radial modes \citet{Soufi95}.
The star studied here is a very rapidly rotating A type star. Therefore 
rotation must be taken
into account, not only when computing equilibrium models but also in the
computation of the oscillation frequencies. The forthcoming space mission 
COROT \citep{Baglin02}, represents a very good opportunity for
investigating such stars since such high frequency resolution data will allow us to 
test theoretically predicted effects of rotation.
\begin{figure*}
  \begin{center}
  \includegraphics[height=10cm]{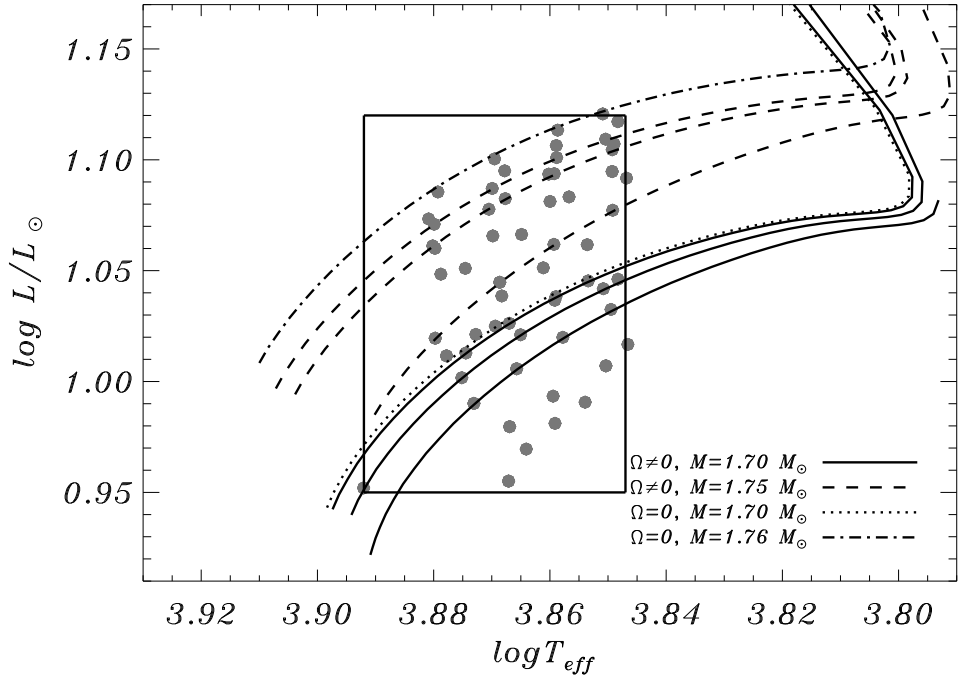}
  \caption{HR diagram containing the whole sample of 
           models considered in this work (filled circles) as well as the
	   observational photometric error box. Dotted and dot-dashed lines
	   correspond to non--rotating evolutionary tracks of $1.70$ and $1.76\,\msol$ 
	   models respectively. Continuous 
	   curves represent those of $1.70\,\msol$ models with
	   rotational velocities, from top to bottom, of $50$, $100$ and
	   $150\,\kms$ respectively. Finally, dashed lines represent evolutionary
	   tracks of $1.75\,\msol$ models with rotational velocities varying, from
	   top to bottom, of $50$, $100$ and $200\,\kms$ respectively.}
  \label{fig:altair_hr}
  \end{center}
\end{figure*}
The paper is structured as follows: In Sect.~\ref{sec:fundparam}, fundamental 
parameters necessary for the modelling of Altair are given. Equilibrium models
as well as our adiabatic oscillation code are described in Sect.~\ref{sec:models}.
Section~\ref{sec:comparison} presents a discussion of the two seismic approaches followed:
1) considering radial modes and analysing the Petersen diagrams, and 2) 
considering radial and non radial modes. We also discuss a possible modal 
identification.
Finally, major problems encountered for modelling Altair and conclusions are
presented in Sect.~\ref{sec:conclusions}.

\section{Fundamental parameters\label{sec:fundparam}}

Several values of the effective temperature and surface gravity of Altair 
($\alpha$~Aql) can be found in the literature. Recently, \citep{ErspamerNorth03} 
proposed 
$\teff\!\!=\!\!7550\,K$ and $\logg=4.13\pm0.3$~dex,
derived from photometric measurements (Geneva system).
From Hipparcos measurements (a parallax of $194.44\pm0.94$ mas) combined with
the observed V magnitude and the bolometric correction given by 
\citet{Flower96}, we obtain a bolometric magnitude of $M_{\mathrm{bol}}=2.18\pm0.1$~dex. 
Using previous values we thus report a luminosity of
$\log(\mathrm{L/L}_{\odot})=0.984\pm0.04$. 

Rapid rotation modifies the location of stars in the HR diagram. 
\citet{MiHer99} proposed a method to determine the effect of rotation
on photometric parameters. In the framework of \dss, this method was then
further developed by \citet{Pe99}, showing errors around $200-300\,K$ and $0.1-0.2$~mag in the 
effective
temperature and absolute magnitude determination respectively \citep[see][ for 
recent results on \dss\ in open clusters]{Sua02aa}.
The error box shown in 
Fig.~\ref{fig:altair_hr} has taken such systematic errors into account and it 
will be the reference for our modelling.
\begin{table*}[htp!]
   \begin{center}
    \caption{Table containing the main characteristics of $M=1.70\,\msol$ computed
    models. From left to right, $\Omega$ represents the rotational velocity
    in $\kms$;
    $\teff$ the effective temperature (on a logarithmic scale); L/$\lsol$ the
    luminosity in solar luminosities (on a logarithmic scale); R/$\rsol$ the stellar radius
    in solar radii; $g$ the surface gravity in cgs (on a logarithmic scale);
    the age in Myr; $\nu(\Pi_0)$ the frequency of
    the fundamental radial mode (in $\muHz$) and finally, $\Pi_1/\Pi_0$ represents the
    ratio between periods of the first overtone and the fundamental radial mode.
    Variables indexed with a $d$ represent the same quantities obtained when including
    near-degeneracy effects.}
    \vspace{1em}
    \renewcommand{\arraystretch}{1.2}
    \begin{tabular}[h!]{ccccccccccc}
    \hline
     & 
$\Omega$ & $\teff$ & L/L$_{\sun}$ & R/$\rsol$ & $g$ & Age & $\nu(\Pi_0)$ & $\Pi_0/\Pi_1$ & $\nu(\Pi_0^d)$ & $(\Pi_0/\Pi_1)^d$
 \\
    \hline
  
m$_1$ & 50 & 3.853 & 1.045 & 2.184 & 3.989 & 1200.0 & 142.553 & 0.772 & 142.504 & 0.772
 \\
  
m$_2$ & 100 & 3.851 & 1.042 & 2.202 & 3.982 & 1192.5 & 140.602 & 0.776 & 139.988 & 0.773
 \\
  
m$_3$ & 150 & 3.849 & 1.033 & 2.192 & 3.986 & 1144.2 & 141.355 & 0.782 & 139.176 & 0.760
 \\
  
m$_4$ & 200 & 3.847 & 1.017 & 2.181 & 3.990 & 1039.0 & 141.711 & 0.792 & 141.711 & 0.757
 \\
  
m$_5$ & 250 & 3.854 & 0.991 & 2.046 & 4.045 & 830.0 & 156.566 & 0.798 & 147.819 & 0.722
 \\
  
m$_6$ & 50 & 3.859 & 1.038 & 2.112 & 4.018 & 1132.0 & 150.315 & 0.773 & 150.166 & 0.772
 \\
  
m$_7$ & 100 & 3.858 & 1.032 & 2.101 & 4.023 & 1094.9 & 151.447 & 0.776 & 150.303 & 0.770
 \\
  
m$_8$ & 150 & 3.858 & 1.020 & 2.079 & 4.032 & 1027.5 & 153.719 & 0.782 & 150.688 & 0.767
 \\
  
m$_9$ & 200 & 3.850 & 1.007 & 2.120 & 4.015 & 959.0 & 148.333 & 0.794 & 141.738 & 0.728
 \\
  
m$_{10}$ & 250 & 3.859 & 0.981 & 1.976 & 4.076 & 741.0 & 165.481 & 0.797 & 156.385 & 0.753
 \\
  
m$_{11}$ & 50 & 3.867 & 1.026 & 2.007 & 4.062 & 1021.0 & 162.571 & 0.773 & 162.616 & 0.773
 \\
  
m$_{12}$ & 100 & 3.865 & 1.021 & 2.013 & 4.060 & 999.0 & 161.760 & 0.776 & 162.960 & 0.785
 \\
  
m$_{13}$ & 150 & 3.866 & 1.006 & 1.972 & 4.078 & 898.3 & 166.858 & 0.782 & 170.549 & 0.799
 \\
  
m$_{14}$ & 200 & 3.859 & 0.993 & 2.001 & 4.065 & 829.0 & 162.627 & 0.792 & 155.581 & 0.758
 \\
  
m$_{15}$ & 250 & 3.864 & 0.970 & 1.906 & 4.107 & 634.0 & 175.117 & 0.797 & 165.244 & 0.752
 \\
  
m$_{16}$ & 50 & 3.874 & 1.013 & 1.910 & 4.106 & 899.4 & 175.432 & 0.773 & 175.445 & 0.773
 \\
  
m$_{17}$ & 100 & 3.875 & 1.002 & 1.879 & 4.119 & 823.0 & 179.763 & 0.776 & 179.988 & 0.784
 \\
  
m$_{18}$ & 150 & 3.873 & 0.990 & 1.872 & 4.123 & 755.7 & 180.692 & 0.782 & 182.011 & 0.805
 \\
  
m$_{19}$ & 200 & 3.867 & 0.980 & 1.903 & 4.108 & 699.0 & 175.909 & 0.791 & 182.864 & 0.822
 \\
  
m$_{20}$ & 250 & 3.867 & 0.955 & 1.848 & 4.134 & 509.0 & 183.512 & 0.798 & 190.824 & 0.885
 \\
    \hline
    \end{tabular}
    \label{tab:altairmodels_m170}
   \end{center}

   \begin{center}
    \caption{Idem Table~\ref{tab:altairmodels_m170} for $M=1.75\msol$
    models.}
    \vspace{1em}
    \renewcommand{\arraystretch}{1.2}
    \begin{tabular}[h!]{ccccccccccc}
    \hline
     & 
$\Omega$ & $\teff$ & L/L$_{\sun}$ & R/$\rsol$ & $g$ & Age & $\nu(\Pi_0)$ & $\Pi_0/\Pi_1$ & $\nu(\Pi_0^d)$ & $(\Pi_0/\Pi_1)^d$
 \\
    \hline
  
m$_{21}$ & 50 & 3.850 & 1.109 & 2.384 & 3.925 & 1211.3 & 126.616 & 0.773 & 126.598 & 0.772
 \\
  
m$_{22}$ & 100 & 3.849 & 1.105 & 2.383 & 3.926 & 1192.5 & 126.544 & 0.776 & 126.205 & 0.771
 \\
  
m$_{23}$ & 150 & 3.849 & 1.095 & 2.356 & 3.935 & 1144.2 & 128.588 & 0.782 & 126.980 & 0.757
 \\
  
m$_{24}$ & 200 & 3.849 & 1.077 & 2.311 & 3.952 & 1040.0 & 132.064 & 0.792 & 134.845 & 0.765
 \\
  
m$_{25}$ & 250 & 3.848 & 1.046 & 2.239 & 3.980 & 820.4 & 137.080 & 0.807 & 147.028 & 0.954
 \\
  
m$_{26}$ & 50 & 3.859 & 1.101 & 2.271 & 3.967 & 1133.8 & 136.626 & 0.773 & 136.584 & 0.773
 \\
  
m$_{27}$ & 100 & 3.859 & 1.094 & 2.248 & 3.976 & 1094.9 & 138.732 & 0.777 & 138.042 & 0.773
 \\
  
m$_{28}$ & 150 & 3.860 & 1.081 & 2.209 & 3.992 & 1027.5 & 142.391 & 0.783 & 139.861 & 0.769
 \\
  
m$_{29}$ & 200 & 3.859 & 1.062 & 2.167 & 4.008 & 911.0 & 146.203 & 0.792 & 140.219 & 0.735
 \\
  
m$_{30}$ & 250 & 3.859 & 1.037 & 2.106 & 4.033 & 727.6 & 151.842 & 0.803 & 151.842 & 0.756
 \\
  
m$_{31}$ & 50 & 3.870 & 1.087 & 2.124 & 4.026 & 1013.7 & 151.508 & 0.773 & 151.667 & 0.774
 \\
  
m$_{32}$ & 100 & 3.868 & 1.082 & 2.135 & 4.021 & 999.0 & 150.273 & 0.777 & 148.854 & 0.769
 \\
  
m$_{33}$ & 150 & 3.870 & 1.066 & 2.073 & 4.047 & 898.2 & 157.023 & 0.783 & 153.231 & 0.764
 \\
  
m$_{34}$ & 200 & 3.869 & 1.045 & 2.035 & 4.063 & 768.6 & 161.177 & 0.792 & 154.113 & 0.758
 \\
  
m$_{35}$ & 250 & 3.869 & 1.025 & 1.982 & 4.086 & 616.2 & 167.403 & 0.799 & 157.393 & 0.751
 \\
  
m$_{36}$ & 50 & 3.880 & 1.071 & 1.991 & 4.082 & 880.3 & 167.222 & 0.773 & 167.240 & 0.773
 \\
  
m$_{37}$ & 100 & 3.880 & 1.061 & 1.967 & 4.093 & 823.0 & 170.376 & 0.776 & 170.733 & 0.783
 \\
  
m$_{38}$ & 150 & 3.879 & 1.048 & 1.951 & 4.100 & 755.7 & 172.435 & 0.782 & 174.374 & 0.807
 \\
  
m$_{39}$ & 200 & 3.880 & 1.020 & 1.878 & 4.132 & 555.0 & 182.397 & 0.792 & 185.953 & 0.848
 \\
  
m$_{40}$ & 250 & 3.878 & 1.012 & 1.878 & 4.133 & 497.4 & 182.263 & 0.796 & 187.153 & 0.866
 \\
    \hline
    \end{tabular}
    \label{tab:altairmodels_m175}
   \end{center}
\end{table*}

\begin{table*}
   \begin{center}
    \caption{Idem Table~\ref{tab:altairmodels_m170} for $M=1.76\msol$
    models. }
    \vspace{1em}
    \renewcommand{\arraystretch}{1.2}
    \begin{tabular}[ht!]{ccccccccccc}
    \hline
   & 
$\Omega$ & $\teff$ & L/L$_{\sun}$ & R/$\rsol$ & $g$ & Age & $\nu(\Pi_0)$ & $\Pi_0/\Pi_1$ & $\nu(\Pi_0^d)$ & $(\Pi_0/\Pi_1)^d$
 \\
    \hline
  
m$_{41}$ & 50 & 3.851 & 1.121 & 2.410 & 3.918 & 1200 & 124.959 & 0.773 & 124.943 & 0.772
 \\
  
m$_{42}$ & 100 & 3.848 & 1.117 & 2.429 & 3.911 & 1192.5 & 123.249 & 0.776 & 122.984 & 0.771
 \\
  
m$_{43}$ & 150 & 3.849 & 1.107 & 2.394 & 3.924 & 1144.2 & 125.873 & 0.782 & 124.407 & 0.756
 \\
  
m$_{44}$ & 200 & 3.847 & 1.092 & 2.375 & 3.931 & 1039.0 & 126.768 & 0.793 & 131.618 & 0.874
 \\
  
m$_{45}$ & 250 & 3.854 & 1.062 & 2.224 & 3.988 & 830.0 & 139.769 & 0.803 & 144.078 & 0.903
 \\
  
m$_{46}$ & 50 & 3.859 & 1.113 & 2.306 & 3.957 & 1132.0 & 133.931 & 0.773 & 133.896 & 0.773
 \\
  
m$_{47}$ & 100 & 3.859 & 1.106 & 2.285 & 3.965 & 1094.9 & 135.737 & 0.777 & 135.124 & 0.773
 \\
  
m$_{48}$ & 150 & 3.860 & 1.094 & 2.238 & 3.983 & 1027.5 & 139.989 & 0.783 & 137.570 & 0.769
 \\
  
m$_{49}$ & 200 & 3.857 & 1.083 & 2.248 & 3.979 & 959.0 & 138.644 & 0.791 & 133.851 & 0.736
 \\
  
m$_{50}$ & 250 & 3.861 & 1.051 & 2.123 & 4.029 & 741.0 & 150.734 & 0.801 & 150.734 & 0.758
 \\
  
m$_{51}$ & 50 & 3.870 & 1.100 & 2.161 & 4.013 & 1021.0 & 148.028 & 0.773 & 147.652 & 0.771
 \\
  
m$_{52}$ & 100 & 3.868 & 1.095 & 2.165 & 4.012 & 999.0 & 147.557 & 0.777 & 146.313 & 0.770
 \\
  
m$_{53}$ & 150 & 3.870 & 1.078 & 2.096 & 4.040 & 898.3 & 154.935 & 0.783 & 151.293 & 0.764
 \\
  
m$_{54}$ & 200 & 3.865 & 1.066 & 2.123 & 4.029 & 829.0 & 151.476 & 0.792 & 145.163 & 0.759
 \\
  
m$_{55}$ & 250 & 3.868 & 1.039 & 2.024 & 4.070 & 634.0 & 162.494 & 0.801 & 152.321 & 0.750
 \\
  
m$_{56}$ & 50 & 3.879 & 1.085 & 2.031 & 4.067 & 899.4 & 162.756 & 0.773 & 162.782 & 0.773
 \\
  
m$_{57}$ & 100 & 3.881 & 1.073 & 1.988 & 4.086 & 823.0 & 168.121 & 0.777 & 168.527 & 0.783
 \\
  
m$_{58}$ & 150 & 3.880 & 1.060 & 1.968 & 4.095 & 755.7 & 170.675 & 0.783 & 172.777 & 0.808
 \\
  
m$_{59}$ & 200 & 3.875 & 1.051 & 1.995 & 4.083 & 699.0 & 166.886 & 0.791 & 159.662 & 0.756
 \\
  
m$_{60}$ & 250 & 3.873 & 1.021 & 1.944 & 4.105 & 509.0 & 172.969 & 0.802 & 161.557 & 0.749
 \\
    \hline
    \end{tabular}
    \label{tab:altairmodels_m176}
   \end{center}
\end{table*}
From photometric measurements, an estimate of the mass ($1.75\pm0.1\msol$) and 
the 
radius ($1.58\,\rsol$) of the star is given by \citet{Zakhozhaj79, Zakhozhaj96}.
However, Altair is found to rotate quite rapidly. Taking advantage of the fact that Altair
is a nearby star, \citet{Richichi02} measured its radius providing 
a diameter of 3.12 mas, which corresponds to a radius of $1.72\,\rsol$. Moreover,
\citet{VanBelle01} showed Altair
as oblate. Their interferometric observations report an equatorial diameter of
 $3.46$ mas (corresponding to a
radius of $1.9\,\rsol$) and a polar diameter of 3.037 mas, for an axial ratio
$a/b=1.14\pm0.029$. The same authors derived a projected rotational velocity of 
$\vsini=210\pm13\,\kms$. Furthermore, \citet{Royer02} reported $\vsini$ values which
range which range from $190\,\kms$ to $250\,\kms$.
In addition to this, recent spectroscopically determined constraints on Altair's inclination 
angle $i$ have been established by \citet{ReinersRoyer04_2}. They provide a range of $i$
between 45$^\circ$ and 68$^\circ$, yielding therefore a range of possible equatorial 
velocities of Altair between $305$ and $245\,\kms$. As shown in next sections, such
velocities, representing 70--90\% of the break up velocity, place significant limits on our ability to model the star.

\section{Modelling the star \label{sec:models}}

\subsection{Equilibrium models\label{ssec:models}}
Stellar models have been computed with the evolutionary code CESAM 
\citep{Morel97}. Around 2000 mesh points for model mesh grid 
(B-splines basis) as well as numerical precision is optimized to compute 
oscillations. 

The equation of state CEFF \citep{ceff} is used, in which, the Coulombian 
correction to the classical EFF \citep{Eggleton73} has been included. 
The $pp$ chain as well as the CNO cycle nuclear reactions are considered,
in which standard species from $^1$H to $^{17}$O are included. The species 
D, $^7$Li and $^7$Be have been set at equilibrium. 
For evolutionary stages considered in this work, a weak electronic screening
in these reactions can be assumed \citep{Clayton68}.
Opacity tables are taken from the OPAL package \citep{Igle96}, complemented at 
low temperatures ($T\leq10^3\,K$) by the tables provided by 
\citep{AlexFergu94}. For the atmosphere reconstruction, the 
Eddington~$T(\tau)$ law (grey approximation) is considered. 
A solar metallicity $Z=0.02$ is used. 
\begin{figure}
 \begin{center}
   \includegraphics[width=8cm]{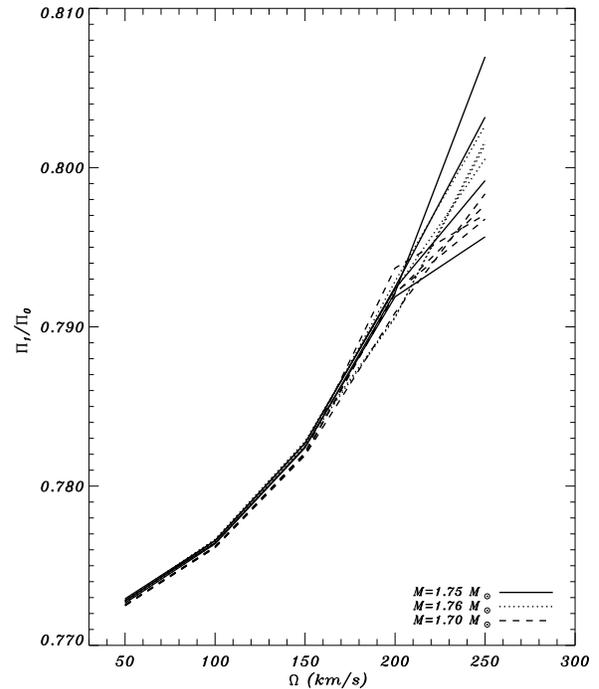}
   \caption{Variation of the first overtone to fundamental radial period ratio
            as a function of the rotational velocity. For each mass considered, four curves are 
	    displayed, representing the results obtained modifying the effective
	    temperature of models from $\teff=3.85$ (bottom) to $\teff=3.88$ (top) in steps
            of 0.01~dex.}   
   \label{fig:p1p0_rot}
 \end{center}
\end{figure}
Convective transport is described by the classical Mixing Length theory,
with efficiency and core overshooting parameters set to
$\alpha_{ML}=l_{m}/\mathrm{H}_p=1.8$ and $d_{ov}=l_{ov}/\mathrm{H}_p=0.2$,
respectively. The latter parameter is prescribed by \citet{Scha92} for
intermediate mass stars. $\mathrm{H}_p$ corresponds to the local pressure 
scale-height, while $l_{m}$ and $d_{ov}$ represent respectively the mixing length 
and the inertial penetration distance of convective elements.

Rotation effects on equilibrium models (\emph{pseudo} rotating models) has been 
considered by modifying the equations \citep{KipWeig90} to include the 
spherically symmetric 
contribution of the centrifugal acceleration by means of an effective gravity
$g_{eff}=g-{\cal A}_c(r)$,
where $g$ corresponds to the local gravity, and ${\cal A}_c(r)$ represents the
radial component of the centrifugal acceleration.
During evolution, models are assumed to rotate as a rigid body, and
their total angular momentum is conserved. 

In order to cover the range of $\vsini$ given in Sect.~\ref{sec:fundparam}, 
a set of rotational velocities of 50, 100, 150, 200 and $250\,\kms$ has been 
considered. The location in the HR diagram of models considered in this work
is given in Fig.~\ref{fig:altair_hr}. 
A wide range of masses and rotational velocities is delimited by the 
photometric error box. To illustrate this, a few representative evolutionary tracks are also 
displayed. Characteristics of computed equilibrium models (filled circles) are
given in Tables~\ref{tab:altairmodels_m170}, \ref{tab:altairmodels_m175} and
\ref{tab:altairmodels_m176}, for models of $1.70\,\msol$, $1.75\,\msol$ and
$1.76\,\msol$ respectively.

\subsection{The oscillation computations\label{ssec:oscil}}

Theoretical oscillation spectra are computed from the equilibrium models described in the previous
section. For this purpose the oscillation code \emph{Filou} \citep{filou,SuaThesis} is used. This
code, based on a perturbative analysis, provides adiabatic oscillations corrected for the effects 
of rotation up to second order (centrifugal and Coriolis forces). 

Furthermore, for moderate--high rotational velocities, the effects of near degeneracy are expected
to be significant \citep{Soufi98}. Two or more modes, close in frequency, are rendered \emph{degenerate} 
by rotation under certain conditions, corresponding to selection rules. In particular these rules select modes
 with
the same azimuthal order $m$ and degrees $\ell$
differing by 2 \citep{Soufi98}. If we consider two generic modes $\alpha_1\equiv(n,\ell,m)$ and
$\alpha_2\equiv(n^\prime,\ell^\prime,m^\prime)$ under the aforementioned conditions, near degeneracy
occurs for $|\sigma_{\alpha_1}-\sigma_{\alpha_2}|\leq\sigma_{\Omega}$,
where $\sigma_{\alpha_1}$ and $\sigma_{\alpha_2}$ represent the eigenfrequency associated to modes
$\alpha_1$ and $\alpha_2$ respectively, and $\sigma_\Omega$ represents the stellar rotational
frequency \citep[see][ for more details]{Goupil00}.
\begin{figure}
 \begin{center}
   \includegraphics[width=8cm]{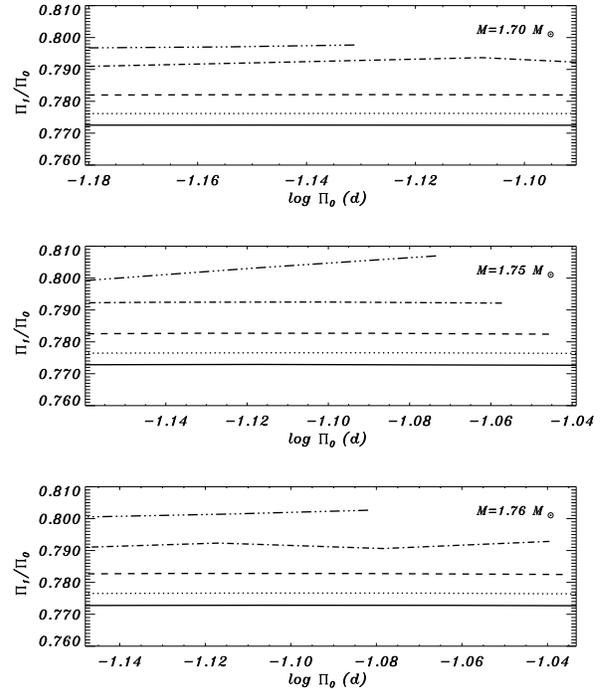}
   \caption{$\Pi_1/\Pi_0$ versus $\log\!\Pi_0$. Three panels are displayed
   corresponding to the three masses considered in this work. For each panel,
   different lines represent the results obtained varying the rotational velocity
   from $\Omega=50\,\kms$ (bottom, solid line) to $\Omega=250\,\kms$ (top,
   dash--dotted line).}
   \label{fig:p1p0_p0}
 \end{center}
\end{figure}

\section{Comparison between theory and observations\label{sec:comparison}}

High-amplitude \dss\ (HADS) display V amplitudes in excess of 0.3~mag and generally oscillate
in radial modes. In contrast, lower-amplitude members of the class present complex spectra,
typically showing non-radial modes. 

The amplitude of the observed main frequency is below 0.5~ppt which is several times smaller 
than typically detected in non-HADS \dss\ from ground-based observations. 
\begin{figure*}
 \begin{center}
   \includegraphics[width=10cm]{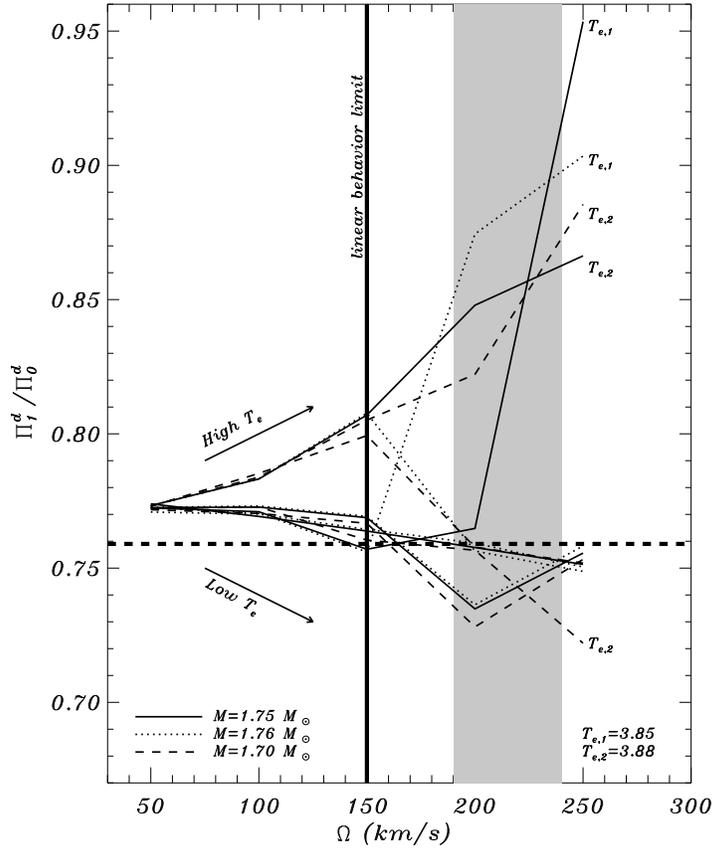}
   \caption{Same as Fig.~\ref{fig:p1p0_rot} but considering near 
            degeneracy. $T_{e,1}$ and $T_{e,2}$ labels represent effective
	    temperatures expressed in a logarithmic scale in K. The vertical 
	    thick solid line indicates the limit for which 
	    $\Pi_1/\Pi_0$ varies linearly with $\Omega$ in presence of near degeneracy (see 
	    details in text). The horizontal dashed thick line represents the 
	    observed $\Pi_1/\Pi_0$. Finally, the shaded area represents the
	    range of rotational velocities $\Omega$ (assuming $i=90$ deg).}
   \label{fig:p1p0_rotc}
 \end{center}
\end{figure*}
As shown in paper I, using the classical period-luminosity relation with the fundamental parameters
given in Sect.~\ref{sec:fundparam}, and assuming the value of $Q = 0.033 \rm~d$ 
given by \citet{Breger79} for \dss, the frequency of the fundamental radial mode is predicted to
be $15.433 \rm~d^{-1}$, suggestively close to the observed $f_1 = 15.768 \rm~d^{-1}$. For the 
remaining modes (Table~\ref{tab:freqobs}) there is no observational evidence to identify them as 
radial. Nevertheless, the second dominant mode, $f_2$, will be considered here as the first 
overtone.

The present study is divided into two parts. In Sect.~\ref{ssec:radialmodes}, we consider
only the two dominant modes (ie. $f_1$ and $f_2$, cf. Table~\ref{tab:freqobs}). The observations 
are then compared with models through period--ratio
vs. period diagrams, from now on called Petersen diagrams 
\citep[see e.g.][]{PetersenDalsgaard96,PetersenDalsgaard96_2}. Then, 
in Sect.~\ref{ssec:nonradialmodes} we use the 5 most dominant observed modes to find the best fit 
to the theoretical models.

\subsection{Considering two radial modes\label{ssec:radialmodes}}

The well known Petersen diagrams show the variation of $\Pi_1/\Pi_0$ ratios with
$\log\Pi_0$, where $\Pi_0$ represents the period of the fundamental radial mode, and
$\Pi_1$ the period of the first overtone. They are quite useful 
for constraining the mass and metallicity of models given the observed ratio between the 
fundamental radial mode and the first overtone. However these diagrams do not consider the effect
of rotation on oscillations 
\citep[see][ for more details]{PetersenDalsgaard96,PetersenDalsgaard96_2}.  
Problems arise when rapid rotation is taken into account: it 
introduces variations in the period ratios even for radial modes. This, combined with the
sensitivity of Petersen diagrams to the metallicity, renders the present 
work somewhat limited. A detailed investigation of such combined effects is currently a work
in progress \citep{Sua05pet1}. 
In order to cover the range of effective temperature (from $\teff=3.85$
to $\teff=3.88$), luminosity and rotational velocity, three sets of 20 models
are computed, where rotational velocity varies from $\Omega=50$ to $\Omega=250\,\kms$.
Characteristics of the three sets are given in Tables \ref{tab:altairmodels_m170},
\ref{tab:altairmodels_m175} and \ref{tab:altairmodels_m176}, corresponding 
to sets of $1.70\msol$, $1.75\msol$ and $1.76\msol$ models respectively.
\begin{figure*}
 \begin{center}
   \includegraphics[width=8cm]{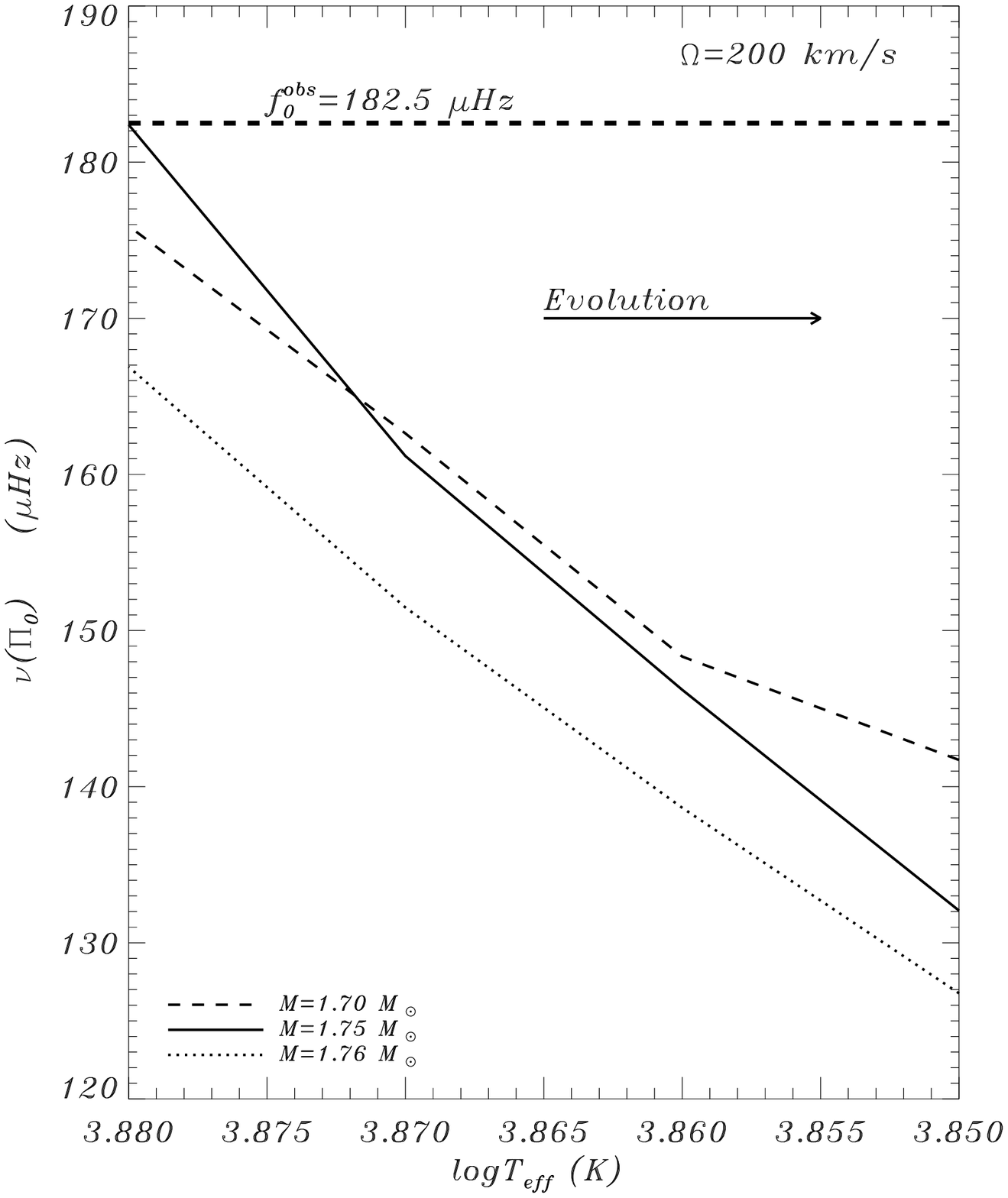}
   \includegraphics[width=8cm]{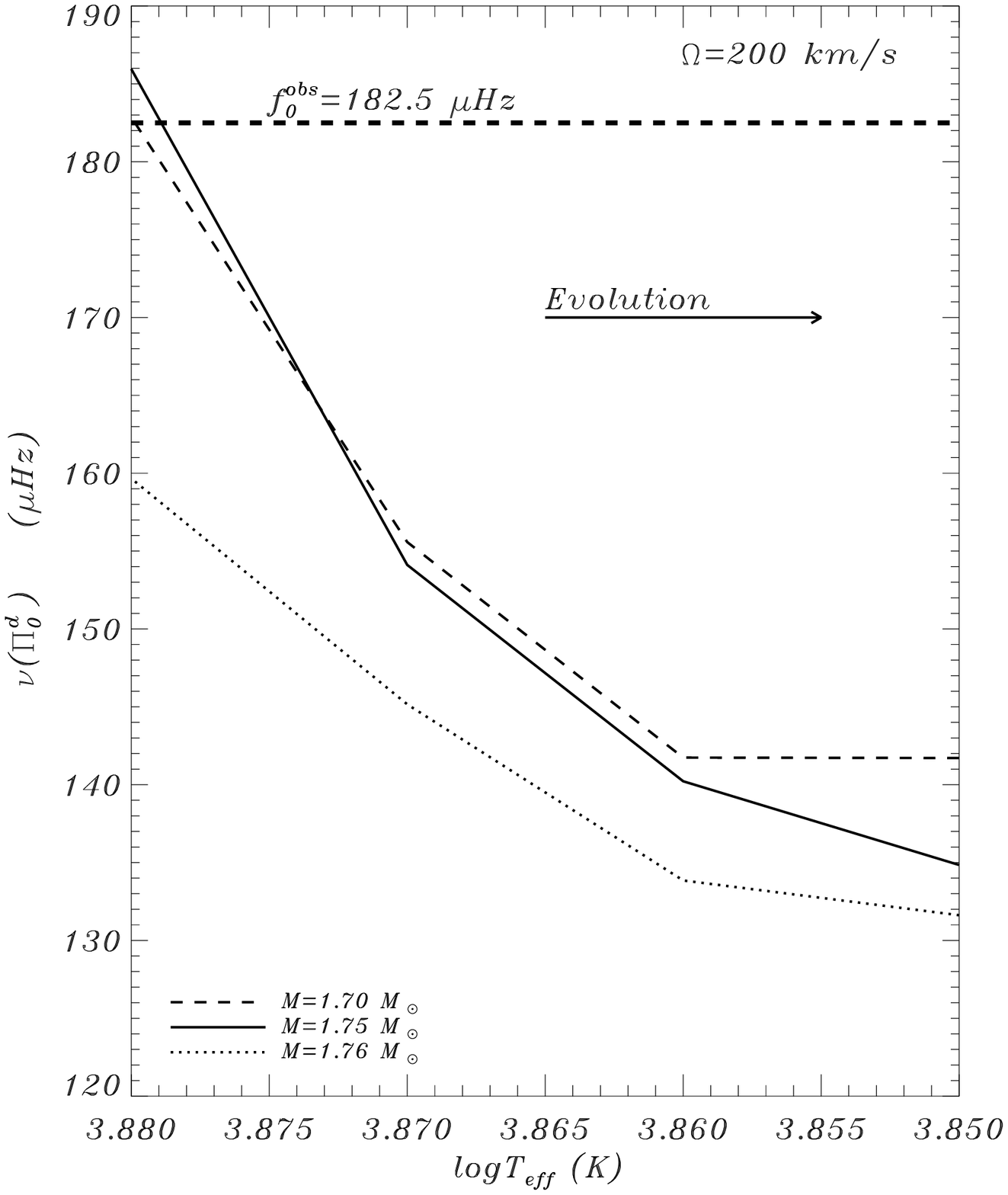}
   \caption{Variation of fundamental mode frequencies $\nu(\Pi_0)$ (in $\muHz$) 
            with effective temperature for models with a rotational velocity
	    of $\Omega=200\,\kms$, representative of Altair. Left panel
	    corresponds to frequencies not corrected for the effect of near
	    degeneracy, and right panel to those corrected. Finally, the 
	    horizontal dashed line represents the frequency of the observed fundamental mode
	    $f_0^{\mathrm{obs}}$.}
   \label{fig:p0_teff}
 \end{center}

\end{figure*}
Adiabatic oscillations are then computed considering radial and non-radial
modes. In order to analyse the Petersen diagrams corresponding to these models
we calculate the $\Pi_1/\Pi_0$ ratios. In Fig.~\ref{fig:p1p0_rot} the $\Pi_1/\Pi_0$ ratios 
are displayed as a function of the
rotational velocity $\Omega$. For each mass (line style) there are four curves
corresponding to four different effective temperatures (from bottom to top). It
can be seen that all ratios increase proportionally with the rotational velocity.
From $\Omega=50\,\kms$ to $\Omega=200\,\kms$ all models behave identically following a
\emph{quasi}--linear relation. For low rotational velocities, 
$\Pi_1/\Pi_0$ ratios approach the canonical 0.77 obtained in absence of rotation.
As expected, dispersion of curves is not significant, since the masses of the models are quite
similar, for a given rotational velocity. However, for $\Omega \geq 200\,\kms$, dispersion of curves also increases as a consequence of errors in the perturbative 
method used for computing
oscillations. With rotational velocities higher than $\sim\!200\,\kms$, results 
should be interpreted carefully, since the second order perturbation theory considered here 
may fail \citep[see][ for more details]{lig01}.
With a projected velocity of $\vsini=190$--$230\,\kms$, Altair is at the limit of
validity.

The impact of these results on Petersen diagrams are shown in Fig.~\ref{fig:p1p0_p0}.
In this diagram the variation of $\Pi_1/\Pi_0$ is displayed versus
$\log\Pi_0$ for the three masses studied. As in Fig.~\ref{fig:p1p0_rot},
note the trend toward the standard value of $\Pi_1/\Pi_0=0.77$
as the rotational velocity decreases. 
Considering the $\Pi_1/\Pi_0$ dependence in isolation, a 
canonical value $\Pi_1/\Pi_0(\Omega)$ can
be established for a given mass. For the range of masses studied
here, the analysis of three panels of Fig.~\ref{fig:p1p0_p0} shows a very 
low dependence on mass.
\begin{table*}
   \begin{center}
    \caption{Table containing a proposed mode identification for the
    observed frequencies given in Table~\ref{tab:freqobs}, using the oscillations
    computed from different models. Values in parentheses correspond to the mode
    identification numbers ($n,\ell,m$). The first five models corresponds to those
    chosen by their proximity to the fundamental radial mode. The last five,
    corresponds to those chosen by minimum $\sigma^2$ value. $\nu_i$ represent
    the theoretical frequencies closest to the observed ones.}
    \vspace{1em}
    \renewcommand{\arraystretch}{1.2}
    \begin{tabular}[ht]{ccccccccccccc}
    \hline
     & 
$\nu_1$ & $\nu_1^d$ & $\nu_2$ & $\nu_2^d$ & $\nu_3$ & $\nu_3^d$ & $\nu_4$ & $\nu_4^d$ & $\nu_5$ & $\nu_5^d$ & $\sigma^2$ & $\sigma_d^2$
 \\
    \hline
  
m$_{18}$ & (-1,2,0) & (1,0,0) & (1,2,2) & (1,2,2) & (2,2,-1) & (2,2,-1) & (-1,2,0) & (-1,2,0) & (-1,2,-1) & (-1,2,-1) & 16.937 & 5.698
 \\
  
m$_{19}$ & (-2,1,0) & (1,0,0) & (3,2,2) & (3,2,2) & (4,0,0) & (4,0,0) & (-3,2,-2) & (-1,1,0) & (-1,1,-1) & (-3,2,-2) & 12.531 & 4.056
 \\
  
m$_{20}$ & (1,0,0) & (-1,2,0) & (1,2,2) & (1,2,2) & (2,2,-1) & (2,2,-1) & (-1,2,0) & (-1,2,0) & (-1,2,-1) & (-1,2,-1) & 4.575 & 4.544
 \\
  
m$_{39}$ & (1,0,0) & (1,1,0) & (2,2,1) & (2,2,1) & (6,2,2) & (2,2,0) & (-2,2,-1) & (-2,2,-1) & (-2,2,-1) & (-2,2,-1) & 21.234 & 7.600
 \\
  
m$_{40}$ & (1,0,0) & (-1,2,-2) & (1,2,0) & (1,2,0) & (3,1,1) & (3,1,1) & (-1,2,-2) & (-1,2,-2) & (1,1,1) & (1,1,1) & 0.763 & 1.242
 \\ 
    \hline
    
m$_{10}$ & (-1,1,0) & (-2,2,0) & (3,2,2) & (3,2,2) & (4,0,0) & (4,0,0) & (-3,2,-2) & (-1,1,0) & (-1,1,-1) & (-3,2,-2) & 1.194 & 0.820
 \\
    
m$_{13}$ & (-1,2,0) & (-1,2,0) & (1,2,2) & (1,2,2) & (2,2,-1) & (2,2,-1) & (-1,2,0) & (-1,2,0) & (-1,2,-1) & (-1,2,-1) & 0.903 & 0.903
 \\
   
m$_{23}$ & (-1,2,-2) & (-1,2,-2) & (3,2,2) & (3,2,2) & (4,1,-1) & (4,1,-1) & (0,2,0) & (0,2,0) & (0,2,0) & (0,2,0) & 0.835 & 0.835
 \\
 
m$_{34}$ & (-1,2,0) & (1,1,0) & (2,2,1) & (2,2,1) & (6,2,2) & (2,2,0) & (-2,2,-1) & (-2,2,-1) & (-2,2,-1) & (-2,2,-1) & 1.698 & 1.228
 \\
 
m$_{58}$ & (-1,2,-2) & (-1,2,-2) & (2,1,1) & (2,1,1) & (3,1,-1) & (3,1,-1) & (-1,1,1) & (-1,1,1) & (1,1,1) & (1,1,1) & 0.639 & 0.940      
 \\
     \hline      
    \end{tabular}
    \label{tab:identif}
   \end{center}
\end{table*}

Up to this point, the effect of near degeneracy has
not been taken into account. Near degeneracy
occurs systematically for close modes (in frequency) following certain selection
rules (see Sect.~\ref{ssec:oscil}). It increases the asymmetry of multiplets
and thereby the behaviour of modes. The higher the value of the rotational 
velocity, the higher the importance of near degeneracy. 
In the present case, for the range of rotational velocities considered, near
degeneracy cannot be neglected. As can be seen in Fig.~\ref{fig:p1p0_rotc},
not all models present the same behaviour with $\Omega$. For rotational velocities
up to $150\,\kms$, a double behaviour is shown, one for lower
effective temperatures ($\Pi_1^d/\Pi_0^d<0.77$), and one for higher effective
temperatures ($\Pi_1^d/\Pi_0^d>0.77$), both remaining linear. This shows the dependence
of $\Pi_1/\Pi_0(\Omega)$ with the evolutionary stage of the star.
However, for higher rotational velocities, particularly on
the right side of the vertical \emph{limit} line, everything becomes confusing. 
 At these rotational velocities, the global effect of rotation up to second 
order (which includes asymmetries and near degeneracy) on multiplets complicates
the use of $\Pi_1^d/\Pi_0^d(\Omega)$ predictions on a Petersen diagram. 
In particular, radial modes are affected by rotation through the distortion
of the star (and thereby its propagation cavity), and through near degeneracy
effects, coupling them with $\ell=2$ modes \citep{Soufi98}.
In addition, other factors such as the 
evolutionary stage and the metallicity must also be taken into account, which
makes the global dependence of Petersen diagrams on rotation rather
complex.

 In Fig.~\ref{fig:p0_teff} (left and right panels), the combined effect of
rotation and evolution on the fundamental radial mode is shown
for representative models with a rotational velocity of $200\,\kms$.
In particular, for the $1.76\,\msol$ model, the stellar radius is 
approximately $10^{-1}\,\rsol$ larger 
than those of $1.70$ and $1.75\,\msol$, which difference is of the order
of $10^{-2}\,\rsol$. As a result, a clear difference of behaviour
between those models is observed. Such a 
difference depends not only on the 
mass, but also on the evolutionary stage, the inital rotational
velocity (at ZAMS), metallicity, etc. In \citet{Sua05gammes},
for a certain small range of masses, 
a similar \emph{non-linear behaviour} is found for predicted unstable mode ranges.
The results of this work may provide a clue to understand this
unsolved question.

The reader should notice that
at these velocities, when frequencies are corrected for near degeneracy effects, their
variations with $\teff$ (and thereby with evolution) are more rapid (right
panel) than was the case for uncorrected ones. 
Moreover, a shift to higher
frequencies is observed when correcting for near degeneracy, which favors the
selection of lower mass objects. 
For low temperatures (i.e. for more evolved models),
\emph{stabilization} is found for low $\nu(\Pi_0^d)$ frequencies, which
can be explained by the joint action of both near degeneracy and 
evolution effects.

 In the present work, following the prescription of \citet{Goupil00}, 
modes are \emph{near degenerate} when their 
proximity in frequency is less or equal to the rotational frequency
of the stellar model ($|\nu_{i}-\nu_j|\lesssim \sigma_{\Omega} $). Considering
all possibilities, 5 models \emph{identify} the observed $f_1$ as the 
fundamental radial mode: m$_{18}$, m$_{19}$, m$_{20}$ ($1.70\,\msol$), m$_{39}$ and
m$_{40}$ ($1.75\,\msol$). For last three models (20, 39 and 40), this happens
without considering near degeneracy effects. For m$_{18}$ and m$_{19}$, the
theoretical fundamental mode approaches $f_1$ when considering near
degeneracy.

In Table~\ref{tab:identif}, radial and non radial identifications (discussed
in the next section) are presented
for selected models. The first set of five models corresponds to those selected
by their identification of $f_1$ as the radial fundamental mode. As can be
seen, no identification is possible when trying to fit the whole set of observed 
frequencies. However $f_3$ is identified as the third overtone by the 
rapid rotating model ($\Omega=200\,\kms$) m$_{19}$.
Uncertainties in the \emph{observed} mass and metallicity are also an important
source of error in determining the correct equilibrium model. Thus, the fact 
of obtaining fundamental modes with frequencies lower than $180\,\muHz$ for most 
of the models could be explained by an erroneous
position of the photometric box on the HR diagram. In fact, the lower the
mass of model used (always within the errors), the higher the $\nu(\Pi_0)$ value.
However, the location in the HR diagram of models with masses lower 
than $1.70\,\msol$ (with the same metallicity) is not representative of 
the Altair observations. 

At this stage, there are two crucial aspects to fix. On one hand, it is necessary
to determine the physical conditions which enforce degeneracy between mode
frequencies. A physically selective near--degeneracy could
explain the behaviour of $\Pi_1/\Pi_0$ for very high rotating stars. 
On the other
hand, for high rotational velocities, third order effects of rotation are
presumably important.

\subsection{Non-radial modes. \label{ssec:nonradialmodes}}

As neither observational nor theoretical evidence supporting a radial identification of the 
observed modes $f_j$ exists, in this part of the work, we carry out an analysis
generalised to non-radial modes. 
As we did for radial oscillations (see Sect.~\ref{ssec:oscil}), here we compute non-radial adiabatic oscillations for the models of
Tables~\ref{tab:altairmodels_m170}, \ref{tab:altairmodels_m175} and \ref{tab:altairmodels_m176}. 
Assuming the 7 observed modes of Altair pulsate with
$\ell\leq3$, a possible mode identification is proposed in Table~\ref{tab:identif}.
In order to avoid confusion, only the first 5 observed modes (Table~\ref{tab:freqobs}) with larger
amplitudes, will be considered.
This constitutes a \emph{rough} identification based only on the proximity
between observed and theoretical mode frequencies for each model. That is, no additional 
information about the $\ell$ and $m$ values is used.

In order to obtain an estimate of the quality of fits (identification of the whole set
of observed frequencies), the mean square error function is used. The lower its value, the 
better the fit for the free parameters considered. 
For each model 
\eqn{\sigma^2=\frac{1}{N}\sum_{i=1}^N\,\Big(f_{o,i}-\nu_{t,i}\Big)^2\label{eq:defchi2}}
is calculated, where $f_{o,i}$ and $\nu_{t,i}$ represent the observed and theoretical frequencies 
respectively.
The total number of observed frequencies is represented by $N=5$. Calculations are made
fixing the metallicity $Z$, the overshooting parameter $d_{\mathrm{ov}}$ and the
mixing length parameter $\alpha_{\mathrm{ML}}$ (see Sect.~\ref{ssec:models}). On the other side,
the mass M, the rotational velocity $\Omega$ and the evolutionary stage $X_c$ have been
considered as free parameters. In Table~\ref{tab:identif}, the last two columns list
$\sigma^2(\mathrm{M},\Omega,X_c)$ calculated for each model, both without 
near degeneracy ($\sigma^2$) and including it ($\sigma_d^2$). When considering the whole set of 
models computed, the inclusion of near degeneracy increases the quality of 
fits by roughly 30\%.
\begin{figure*}
  \begin{center}
  \includegraphics[height=10cm]{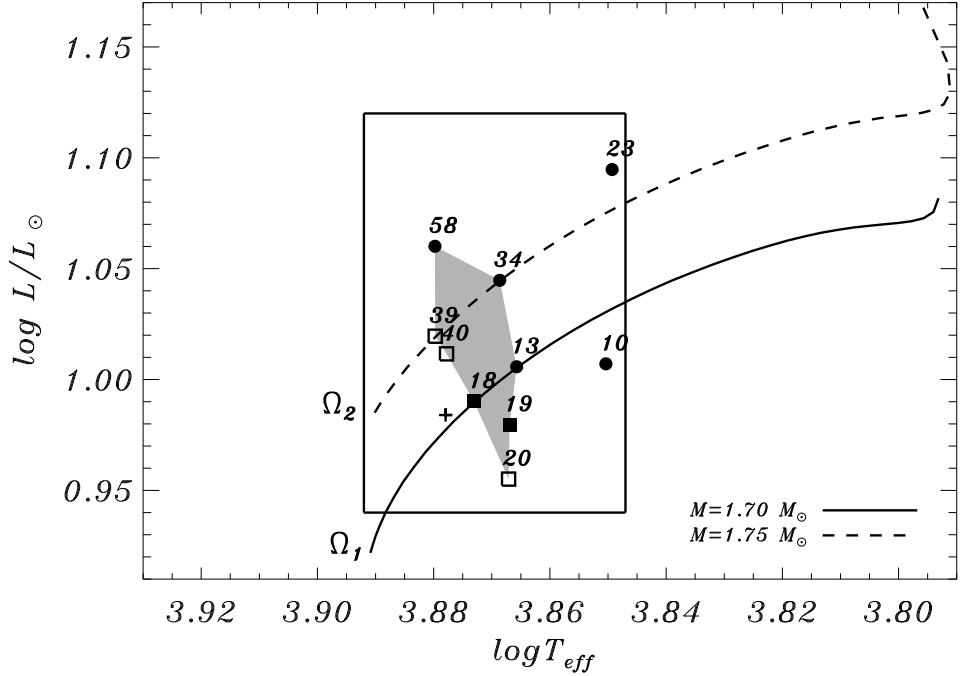}
  \caption{HR diagram showing selected models given in Table~\ref{tab:identif}. 
           Labels correspond to model numbers.
           Square symbols represent models selected by the proximity to the fundamental
	   radial mode; filled squares are used for those taking into account near degeneracy
	   effects. Filled circles represent models selected by the minimum $\sigma^2$ value.
	    As a reference, 
	   observations (the cross symbol) and
	   two evolutionary tracks are depicted: One corresponding to a $1.70\,\msol$
	   and $\Omega_1=150\,\kms$ (solid line), and the other one, corresponding to
	   a $1.75\,\msol$ and $\Omega_2=200\,\kms$ (dashed line). For more details 
	   see the text.}
  \label{fig:altair_hr_sel}
  \end{center}
\end{figure*}
Table~\ref{tab:identif} is divided in two parts: the first 5 models have been chosen as
identifying $f_1$ as the fundamental radial mode. The second set of models correspond
to those with minimizing $\sigma^2$. Analysing the results obtained for the 14 
selected models it can be seen that no $\ell=3$ identifications are found and 
most of identifications correspond to $\ell=2$ (~60\%) and $\ell=1$ (~32\%) modes. 
As happened for the first set of models, $f_3$ is identified as the third radial overtone by 
models m$_{10}$. 

On the other hand, since no specific clues for the degree $\ell$ and the azimuthal order $m$
are given, other characteristics of the observed spectrum must be investigated. In particular,
it is quite reasonable to consider some of the observed frequencies as belonging to one
or more rotational multiplets. Although no complete multiplets are found, several sets
of two multiplet members are obtained. Specifically, the observed frequencies $f_4$ and 
$f_5$ are identified as members of $\ell=2$ multiplets by models m$_{13}$, m$_{18}$, m$_{20}$, 
m$_{39}$ and m$_{10}$, and as members of $\ell=1$ triplets by models m$_{10}$ 
and m$_{19}$.

Figure~\ref{fig:altair_hr_sel} shows the location on 
a HR diagram of selected models given in Table~\ref{tab:identif}. Notice that around 80\% of them 
lie in effective temperature and luminosity ranges of $\log\teff=[3.88,3.85]$ and 
$\log\mathrm{L/}\lsol=[0.96,1.09]$ respectively. In the area delimited by these models (shaded surface),
models with similar characteristics are found. As can be seen, the shaded area is located in
the central part of the error box. Except models m$_{39}$, m$_{40}$ and m$_{58}$, all
models are situated toward colder and more luminous locations in the box. This is in agreement 
with the expected effect of rotation on fundamental parameters (see Sect.~\ref{sec:fundparam}).
On the other hand, positions of models selected by the proximity to the fundamental
radial mode (squares) can be connected by a straight line. This is an iso-${\bar \rho}$ 
line\footnote{Line which connects models with the same stellar mean density.}, with
${\bar \rho}\simeq5.23\,\rm{g}\,\rm{cm}^3$, and
which basically explains the similar frequencies of their fundamental
radial mode. Considering near-degeneracy, the identification of the fundamental radial mode
and the lowest $\sigma^2$ value, our \emph{best} model is m$_{19}$, which is at the lower 
limit of the range of Altair's observed $\vsini$. 

Nevertheless, in order to further constrain our models representative of Altair, it would be 
necessary
to obtain additional information on the mode degree and/or azimuthal order of observations.
In this context, spectroscopic analysis may provide information about $\ell$, $m$ and the angle
of inclination of the star, as nonradial pulsations generate Doppler shifts and line profile
variations \citep{AertsEyer00}. Furthermore, multicolor photometry may also provide information
about $\ell$ \citep{Garrido90}. 


\section{Conclusions\label{sec:conclusions}}

In the present paper a theoretical analysis of frequencies of \object{HR6534} (Altair) 
was presented,
where rapid rotation has been properly taken into account in the modelling. 
The analysis was separated in two parts: 1) considering the observed modes $f_1$ and $f_2$ 
corresponding to the fundamental radial mode and the first overtone and models were analysed
through the Petersen diagrams, and 2) a preliminary modal identification was proposed by
considering radial and non-radial oscillations. 

Firstly, in the context of radial modes, we studied the isolated effect of rotation on
Petersen diagrams. For the different rotational velocities
considered, the shape of $\Pi_1/\Pi_0(\Omega)$ leads to a limit of 
validity of the perturbation theory (up to second order) used at around $\Omega=200\,\kms$.
This limit is mainly given by the behaviour of such period ratios when near degeneracy
is considered, which visibly complicates the interpretation of Petersen diagrams.
Nevertheless, in this procedure, for rotational velocities up to $150\,\kms$ it is
found that $\Pi_1/\Pi_0$ is lower than 0.77 for lower effective temperatures,
and reciprocally higher than 0.77 for higher effective temperatures (inside
the photometric error box). 
The analysis of radial modes also reveals that only a few models identify the 
observed $f_1$ as the 
fundamental radial mode. This reduces the sample of models to those with
masses around $1.70\,\msol$ or lower, on the main sequence and implies models with 
different metallicity in order to agree with photometric error box.
These partial results (for a given region in the HR diagram,
for a given range of masses, evolutionary stages and a given metallicity)
constitute a promising tool for seismic investigation, not only of \dss, but 
in general, of multi-periodic rotating stars. A detailed 
investigation on the
effect of both rotation and metallicity on Petersen diagrams will be proposed in
a coming paper \citep{Sua05pet1}.

Secondly, in the context of radial and non radial modes and assuming observed modes pulsating 
with $\ell=0,1,2,3$, a set of 14 models was selected in which five of them 
identify the fundamental radial order and at least six others identify two of the observed
frequencies ($f_4$ and $f_5$) as members of $\ell=2$ and $\ell=1$ multiplets. A range of masses of 
[1.70,1.76] $\msol$ principally for a wide range of evolutionary stages on the main sequence
(ages from 225 to 775 Myr) was obtained. Considering information of both radial 
(through Petersen diagrams) and non-radial modes, a set of representative models with rotational 
velocities larger than $150\,\kms$ was obtained. 	

Further constraints on the models are thus necessary. Such constraints can be obtained by employing
additional information on the mode degree $\ell$ and/or the azimuthal order $m$ of the observed
modes, which may be inferred from spectroscopy and multicolor photometry 
\citep{Garrido90}. Improvements on adiabatic oscillation computations (including third
order computations) will constitute a coherent and very powerful tool to obtain seismic data 
from future space missions like COROT, EDDINGTON and MOST.

\acknowledgements{This work was partially financed by the Spanish Plan Nacional del Espacio, under 
                  project ESP2004-03855-C03-03, and by the Spanish Plan Nacional de Astronom\'{\i}a y 
		  Astrof\'{\i}sica, under proyect AYA2003-04651. We also thank the anonymous referee
		  for useful comments and corrections which helped us to improve this
		  manuscript.}

\bibliography{2410ref}
\bibliographystyle{aa}

 \end{document}